\documentclass{svjour3}                     
\smartqed  

\usepackage[square,sort,comma,numbers]{natbib}
\usepackage{graphicx}
\usepackage{tikz}
\usetikzlibrary{shapes,arrows}
\usepackage{verbatim}
\usepackage{color}
\usepackage{amsmath}

\renewcommand{\Phi}{D_j}

\begin{document}

\title{Robust and fast online identification of streamwise vortices properties for closed-loop control purposes}

\author{Caroline Braud$^{1}$ \and Alex Liberzon$^{2}$}

\institute{C. Braud \at
              LHEEA Laboratory (ECN/CNRS), 1 rue de la Noë, 44321 Nantes, France \\
              Tel.: +123-45-678910\\
              Fax: +123-45-678910\\
              \email{caroline.braud@ec-nantes.fr}           
           \and
           A. Liberzon \at
              School of Mechanical Engineering, Tel Aviv University, 69978 Tel Aviv, Israel
}

\maketitle

\abstract{
We propose to combine the active vortex generators with the particle image velocimetry (PIV) measurements and post-processing streamwise vortex characterization algorithms into a feedback based closed-loop control system for wind turbine applications. We develop two vortex identification and characterization methods that use PIV realizations for the purpose of a real-time (online or on-the-fly) feedback-based control. Both methods can extract centers and strengths of streamwise vortices generated behind active vortex generators in a turbulent boundary layer flow, and we show how to integrate those in a closed-loop control strategy. For demonstration purposes we use stereoscopic PIV measurements at the wind tunnel facility obtained in the transverse-wall-normal plane behind active vortex generators. A robust algorithm is using the $Q$-criteria and the integration of vorticity of each extracted vortex. Results show that a moving window average of a small number of instantaneous fields is nevertheless needed for increased robustness. The robust method requires the full field PIV computation followed by spatial derivatives calculations. A faster method is developed, which, using only horizontal lines of vertical velocity, has a high potential to significantly cut down the computational effort relative to the robust method. We compare the two methods and discuss their shortcomings and the potential for the real-time, online, closed-loop control of turbulent boundary layers of the wind turbine blades.  
}


\section{Introduction}

Vortex generators (VGs) -- devices generating streamwise vortices which bring high momentum fluid near the surface so that it delays the apparition of flow separation -- are widely used to control separation in turbulent boundary layers, as reviewed by Lin~\cite{lin2002}. Active devices that reproduce streamwise vortices by blowing jets from the wall, rapidly replace the passive VGs, mainly because of the high potential for use in feedback based control loops \cite{shaqarin2013}. High shear from the jet edges induces a roll-up of the boundary layer flow \cite{peterson2004} and this jet-free stream interaction leads to a pair of counter-rotating streamwise vortices generated behind a single jet blown perpendicular to the wall. When the jet has an angle to the wall (pitch and/or skew angle), two counter-rotating vortices are initially created downstream of the device and evolve rapidly into a single coherent vortex of one sign accompanied by a much smaller and weaker region of circulation of the opposite sign near the wall \cite{tilmann2000}. Properties of the generated vortex, such as circulation, distance to the wall and size, among others, dictate the efficiency of the boundary layer separation control.

Sensors used in feedback loops, such as pressure sensors \cite{becker2007}, hot-film sensors \cite{shaqarin2013} and similar, are typically installed at the wall. However, sensing of flow properties at the wall is not adapted to detect vortex properties which are of crucial importance for the boundary layer control. Sensing methods based on flow measurements above the wall using particle image velocimetry (PIV)  have been recently demonstrated for closed-loop control using low Reynolds benchmarks~\cite{gautier_aider2013, willert_al2010}. The control objective in these studies was to re-attach the flow using the vortex structures from the Kelvin-Helmholtz instability of the separated flow region. However, once attached, it is impossible to control the flow until the flow separates again and new vortex structures appear in the flow.

We propose to combine the jet-based active VGs with the real-time PIV sensing tuned to measure the vortices due to active devices. The goal is to maintain the control action under external perturbation while the flow is attached. Our control strategy is built upon the concept of real time PIV sensing (i.e. acquisition, processing and post-processing) that extracts properties of streamwise vortices produced behind the control jets/free-stream interaction near active VGs. Specifically the wind turbine applications are targeted and the control needs to be maintained under atmospheric perturbations.

The targeted control strategy is explained in the following Section~\ref{sec:control_strategy}. The experiment and experimental database acquired and used in this paper are described in the Section~\ref{sec:database}. Two vortex characterization algorithms, a robust and a fast (real-time), that identify the generated streamwise vortices and extract their properties, are presented in Sections~\ref{sec:robust_detect} and \ref{sec:fast_detect}, respectively. Finally, we discuss a possible control strategy implementation using these methods, followed by conclusions in Section~\ref{sec:conclusions}.

\section{Control strategy}
\label{sec:control_strategy}

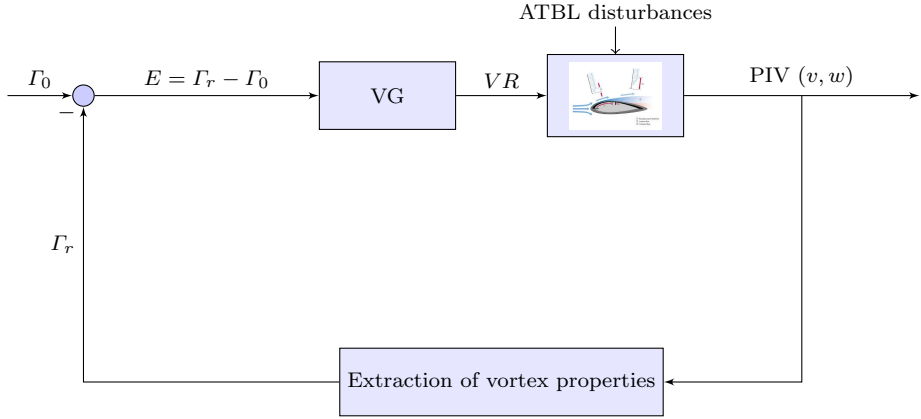
\begin{figure}[htbp]
\begin{center}
\tikzstyle{block} = [draw, fill=blue!10, rectangle,
    minimum height=3em, minimum width=6em]
\tikzstyle{sum} = [draw, fill=blue!20, circle, node distance=1cm]
\tikzstyle{input} = [coordinate]
\tikzstyle{output} = [coordinate]
\tikzstyle{pinstyle} = [pin edge={to-,thin,black}]

\begin{tikzpicture}[auto, node distance=4cm,>=latex']
    \node [input, name=input] {};
    \node [sum, right of=input] (sum) {};
    \node [block, right of=sum] (controller) {VG};
    \node [block, right of=controller, pin={[pinstyle]above:ATBL disturbances},
            node distance=3cm] (system) {\includegraphics[width=0.1\textwidth]{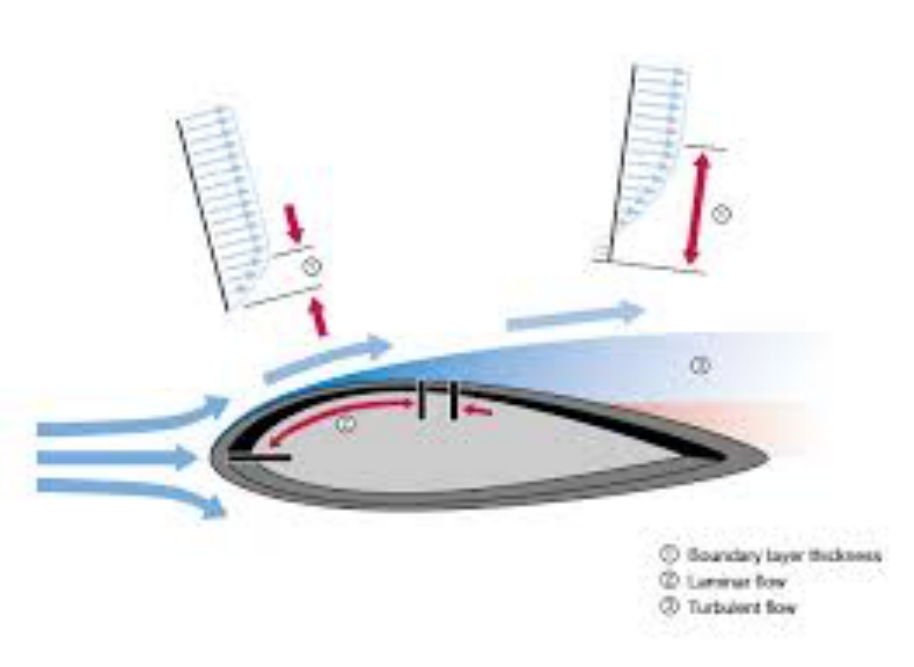}};
    \draw [->] (controller) -- node[name=u] {$VR$} (system);
    \node [output, right of=system] (output) {};
    \node [block, below of=u] (measurements) {Extraction of vortex properties};

    \draw [draw,->] (input) -- node {$\Gamma_0$} (sum);
    \draw [->] (sum) -- node {$E=\Gamma_r-\Gamma_0$} (controller);
    \draw [->] (system) -- node [name=y] {PIV ($v, w$)}(output);
    \draw [->] (y) |- (measurements);
    \draw [->] (measurements) -| node[pos=0.99] {$-$}
        node [near end] {$\Gamma_r$} (sum);
\label{feebackloop}
\end{tikzpicture}
\caption{Block diagram of the targeted control strategy. $\Gamma_0$ is the desired circulation of the streamwise vortex. $E=\Gamma_r-\Gamma_0$ is the error value that is maintained as small as possible using the control variable, $VR$,  which drives to the jet exit amplitude and thus the streamwise vortex properties. A 2D PIV image containing the in-plane component of the velocity field in the spanwise wall-normal plane, $(v,w)$, is the measured output. $\Gamma_r$ is the reduced output which represent the strength of the vortex, extracted from the PIV field using the algorithms described in Section~\ref{sec:fast_detect}.}
\label{cmpd:control}
\end{center}
\end{figure}

In wind turbine applications, that we target with this approach, disturbances arrive from various sources, such as a) variations of speed during blade rotation due to the boundary layer gradient, b) due to misalignment of the rotor relatively to the wind direction, and c) from turbulence from the unsteady atmosphere. Therefore, a lot of external perturbations can prevent the control efficiency. It takes approximately 1 second for the rotor blade to do one loop and therefore to cross the atmospheric boundary layer gradient. If we target to maintain the control action during a rotation of the rotor, the strategy could be the one described in the block diagram in figure~\ref{cmpd:control} using PIV images every 0.01 second (100 Hz).

This strategy uses properties of the streamwise vortices produced downstream active VGs to maintain the lift value whatever the perturbation from the vertical gradient of the atmosphere.  A simplest control goal can be to maintain the vortex circulation constant as described in the diagram in figure~\ref{cmpd:control}. Variations of the blade speed within the gradient of wind speed in the boundary layer will locally decrease or increase the inlet velocity, which is equivalent to the effect of modifying the shear with the controlled jet amplitude and thus decreasing velocity ratio $VR=V_j/U_\infty$, $V_j$ is the jet exit velocity and $U_\infty$ is the wind speed. Under the hypothesis that the Reynolds number is not varying significantly for velocities involved in the gradient of the atmospheric boundary layer, modifying $VR$ is equivalent to modify the inlet velocity. We first apply the control to achieve the desired lift (calibrated off-line), that is maintained during the rotation of the blade using modification of $VR$ according to a steady state law. 

\section{Database}\label{sec:database}

We use stereoscopic PIV (SPIV) experimental dataset obtained in the LML wind tunnel facility, behind active vortex generators (AVGs) embedded in a highly turbulent boundary layer. Details on the wind tunnel facility, control set-up and SPIV measurements can be found in \cite{carlier2005, foucaut_al2014} and only briefly repeated below for the sake of completeness. Use of the stereoscopic setup is important for the accurate measurement of the streamwise vorticity in the spanwise - wall-normal plane. 

\subsection{Wind tunnel}\label{sec:wind_tunnel}

The turbulent boundary layer which develops in the LML wind tunnel is extensively described and characterized in \cite{carlier2005}. The working section is 1 m high, 2m wide and 21.6 m long. The boundary layer thickness that develops in the test section is  $\delta=0.3$ m at the free-stream velocity $U_\infty=8.5$ m/s and do not change much throughout the test section. The turbulence level in the free stream is about $0.3\%$ of $U_\infty$ , and the temperature is kept within $±0.2C^\circ$ by use of an air-water heat exchanger in the plenum chamber. 

\subsection{Active vortex generators}\label{sec:active_vg}

A transverse line of active devices was set in the test section by drilling holes at the wall with an angle to the wall of $45^\circ$ and an angle to the free stream of $45^\circ$ as schematically shown in figure~\ref{fig:SPIV_setup}. Active devices were arranged in a counter-rotating configuration (the angles to the wall are $45^\circ$ and $-45^\circ$ for devices composing the pair), so that three pairs can fit in the transverse direction of the section test. The distance between devices within a counter-rotating pair is fixed to $15\Phi$ where $\Phi=10$ mm is the jet diameter. The distance between pairs is $33.6\Phi$. Three jet velocity ratios were tested, $VR=V_j/U_\infty=3,4,5$ with $V_j$ the jet exit velocity. The momentum coefficient is given by $C_\mu=Q_mV_j/(P_\infty S)$ with $Q_m$ the mass flow rate, $P_\infty$ the dynamic pressure, $S$ the total surface that is controlled (including areas between jets). For incompressible controlled jets, which is the case in the present experiments, this is proportional to $VR$: 
\begin{equation}
C_\mu=\frac{2S_j}{S}VR^2
\end{equation}
\noindent with $S_j=n\pi \Phi^2$ the total surface of controlled jets with $n=4$ the number of jets. The ratio $VR$ will thus be used instead of $C_{\mu}$ in the following. The air circuit which was used to provide compressed air to the actuators is composed of a compressor, a filtration system, a proportional valve, two volumetric flow meters, a manometer, a thermometer and a large reservoir of 90 liters. More details on the air circuit arrangement can be found in Ref.~\cite{shaqarin2013}.

\subsection{SPIV measurements}\label{sec:spiv}

\begin{figure}[htbp]
\begin{center}
\includegraphics[width=0.7\textwidth]{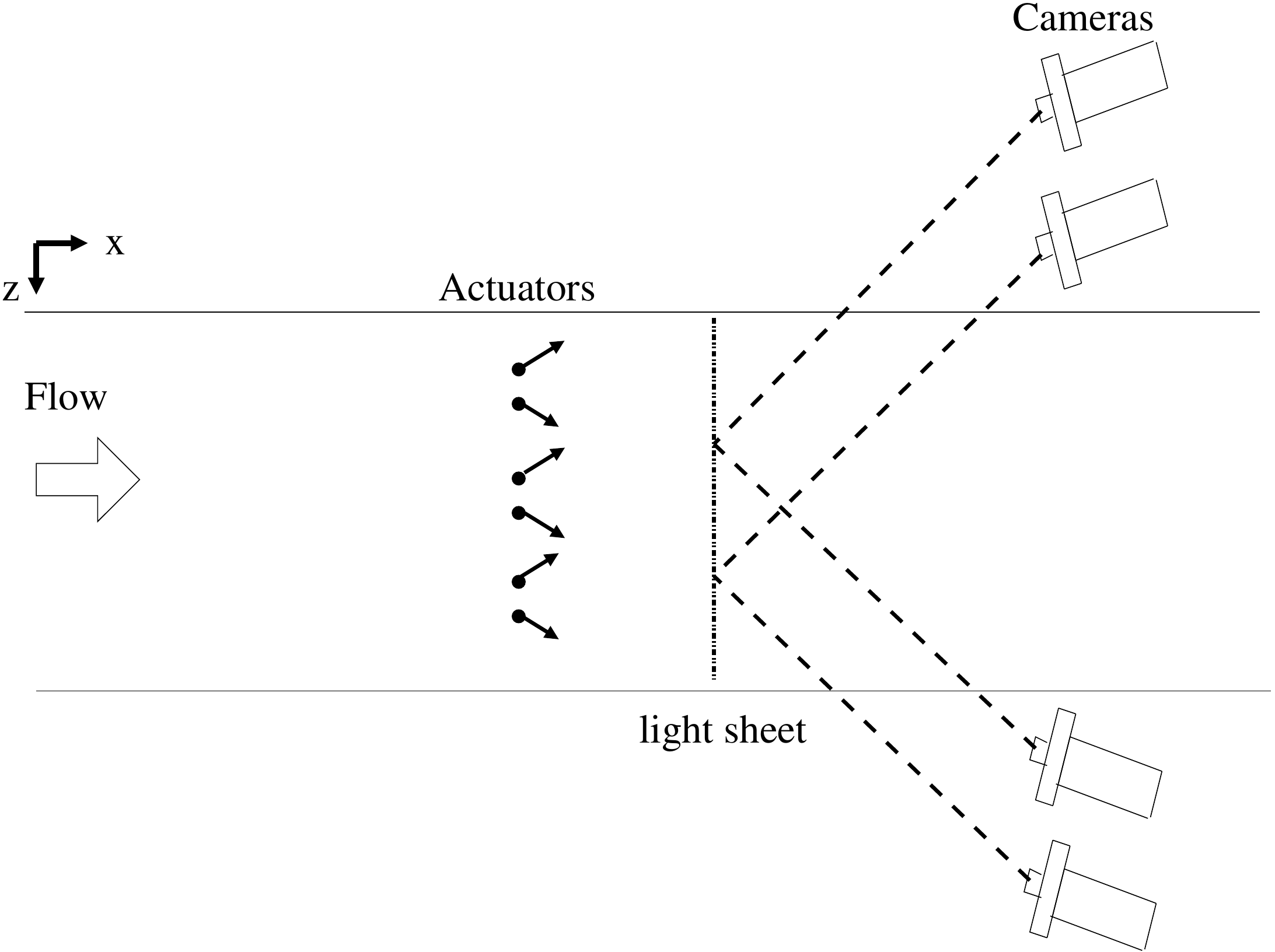} 
\caption{Stereoscopic PIV set-up. Arrows mark the positions and directions of the vortex actuator jets. Two SPIV systems were used to increase the field of view and resolution of the measurements.}
\label{fig:SPIV_setup}
\end{center}
\end{figure}

Two stereo PIV  planes were acquired simultaneously at $x=72\Phi$ from the active devices to obtain a large measurement field  (see figure \ref{fig:SPIV_setup}). This allows us to acquire a large field of view of $30\times70$ cm$^2$, with a resolution down to 2.3 mm (or $0.0076\delta$). This was a compromise to get, at least, a full pair of counter-rotating devices with a sufficient resolution to describe streamwise vortices produced by control devices.  The framework of PIV images is as follows: $x$ is the streamwise direction, perpendicular to the PIV field plane, y is the wall normal direction and z is the transverse direction ($z$=0 at the middle of the PIV field, at the center axis of the counter-rotating vortex pair). The SPIV provides all the velocity components in the measurement plane. However, only the in-plane vector fields ($v, w$) that contain a cross-section of streamwise vortices produced by active devices are used in the analysis and for the control purposes.


\section{Vortex identification methods}\label{sec:robust_detect}

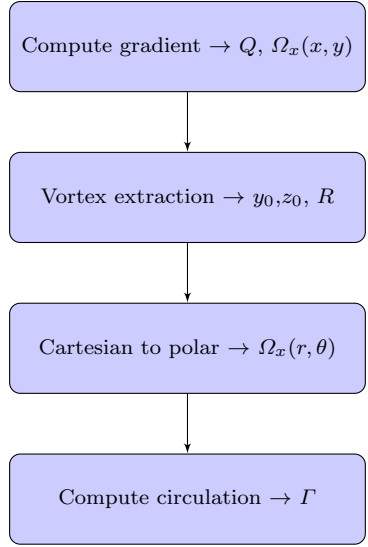
\begin{figure}[ht]
\begin{center}
\tikzstyle{decision} = [diamond, draw, fill=blue!20,
    text width=4.5em, text badly centered, node distance=3cm, inner sep=0pt]
\tikzstyle{block} = [rectangle, draw, fill=blue!20,
    text width=15em, text centered, rounded corners, minimum height=4em]
\tikzstyle{line} = [draw, -latex']
\tikzstyle{cloud} = [draw, ellipse,fill=red!20, node distance=3cm,
    minimum height=2em]
\begin{tikzpicture}[node distance = 2cm, auto]
    \node [block] (gradient) {Compute gradient $\rightarrow$ $Q$, $\Omega_x(x,y)$};
    \node [block, below of=gradient] (extract) {Vortex extraction $\rightarrow$ $y_0$,$z_0$, $R$};
    \node [block, below of=extract] (cart2pol) {Cartesian to polar $\rightarrow$ $\Omega_x(r,\theta)$};
    \node [block, below of=cart2pol] (circulation) {Compute circulation $\rightarrow$ $\Gamma$};
    \path [line] (gradient) -- (extract);
    \path [line] (extract) -- (cart2pol);
    \path [line] (cart2pol) -- (circulation);
\end{tikzpicture}
\caption{Diagram of the algorithm developed to extract the vortex properties (localization $(y_0,z_0)$, radius $R$ and circulation $\Gamma$), of produced vortices behind active vortex generators.}
\label{cmpd:acid}
\end{center}
\end{figure}

Our main goal is to develop methods that will enable real-time or ``online'' streamwise vortex identification and characterization of its properties in a turbulent boundary layer flow. We first developed concepts using an average of 100 SPIV realizations in which vortices are clearly seen and the signal-to-noise ratio is negligible. Then we modify the method to be able to detect vortices and define the properties in instantaneous SPIV realizations. The important parameters of each vortex that will be used for control are a) position of its center, $y_0,z_0$ b) size, $R$, c) circulation, $\Gamma$. These properties can be extracted from a PIV realization using steps described in the diagram in figure~\ref{cmpd:acid}, based on vorticity and the so-called $Q$-criteria \cite{Chakraborty_al2005}.

Gradients in transverse and wall-normal ($y,z$) directions are computed from the two-dimensional PIV realizations from which the longitudinal vorticity $\Omega_x(x,y)$, the rate-of-strain tensor and the $Q$-criterion (without $\partial(.)/\partial x$, $(.)=u,v,w$) are obtained: 
\begin{equation}\label{eq:Q}
Q=\frac{1}{2}\left[||Rot_{ij}||^2-||Str_{ij}||^2\right]
\end{equation}
\noindent with $||Rot_{ij}||=\frac{1}{2}(\partial u_i/\partial x_j + \partial u_j/\partial x_i)$ the rotation rate tensor and $||Str_{ij}||=\frac{1}{2}(\partial u_i/\partial x_j - \partial u_j/\partial x_i)$ the strain rate tensor. First, the vortex center ($y_0$,$z_0)$ is localized using the maximum of $Q$, and radius $R$ is evaluated using  the maximal distance between two neighbor zero-crossing points of the vorticity. Note that due to the use of $Q$ in two-dimensional approximation only, the vortex center position requires an additional, refining, iteration.

\begin{figure}[htbp]
\begin{center}
\includegraphics[width=1.\textwidth]{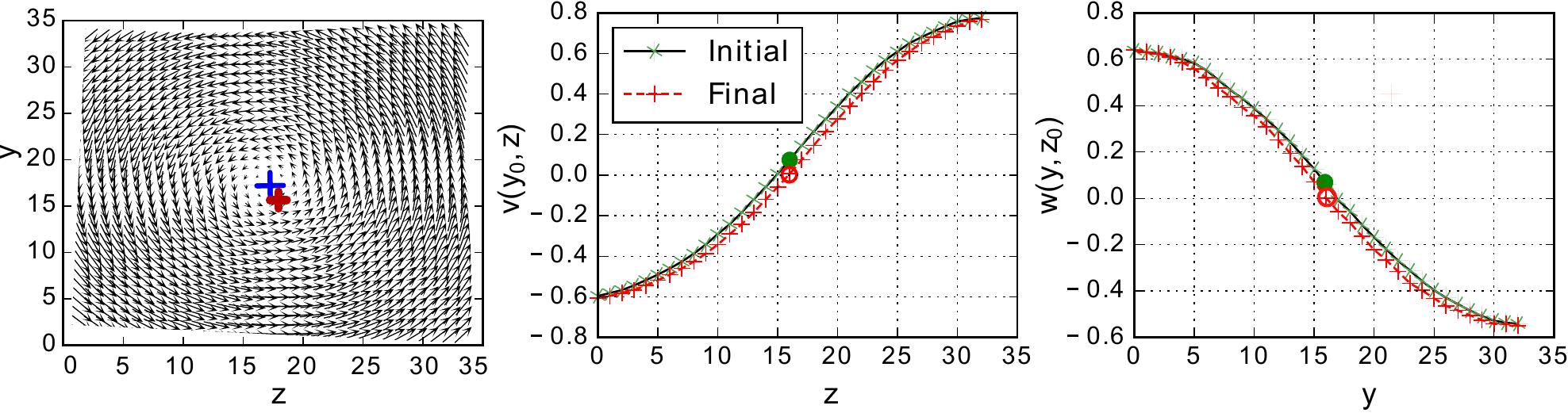} 
 \caption{Refinement procedure on the localization of the vortex center $(y_0,z_0)$, on the in-plane vector field marked by a cross (top), on the vertical velocity component at the vortex center $v(y_0,z)$ marked by a blue round (left) and on the horizontal velocity component at the vortex center $w(y,z_0)$ marked by a blue round (right). Initial and final localization of the vortex center in the refinement procedure are indicated by full and empty symbols, respectively.}
\label{fig:refinement}
\end{center}
\end{figure}

In the presence of a vortex, like those shown in Fig.~\ref{fig:refinement}, an horizontal line of the velocity field at the vortex center will contain mainly the vertical velocity $v(y_0,z)$ component while the horizontal velocity component is close to zero, i.e. $w(y_0,z)\simeq 0$. Similarly, a vertical line of the velocity field at the vortex center will contain mainly the horizontal velocity component $w(y,z_0)$ while the vertical velocity component is close to zero, i.e. $v(y,z_0)\simeq 0$. The vortex center position can be refined searching the position where velocity profiles cross zero lines,  $v(y,z_0)=0$ and $w(y_0,z)=0$. Figure~\ref{fig:refinement} shows an example of this refinement procedure. A single-component velocity profile in horizontal or vertical direction is used and the initial position of the vortex from the rough estimate using $Q$ criterion is marked. Then a nearest zero-crossing point is found and the vortex center position is adjusted. A new radius, $R$, is extracted using the maximal distance between two neighbor maxima of $v$ in $z$ direction (i.e. where $\partial v/\partial z=0$). 

\begin{figure}[htbp]
\begin{center}
\includegraphics[width=0.9\textwidth]{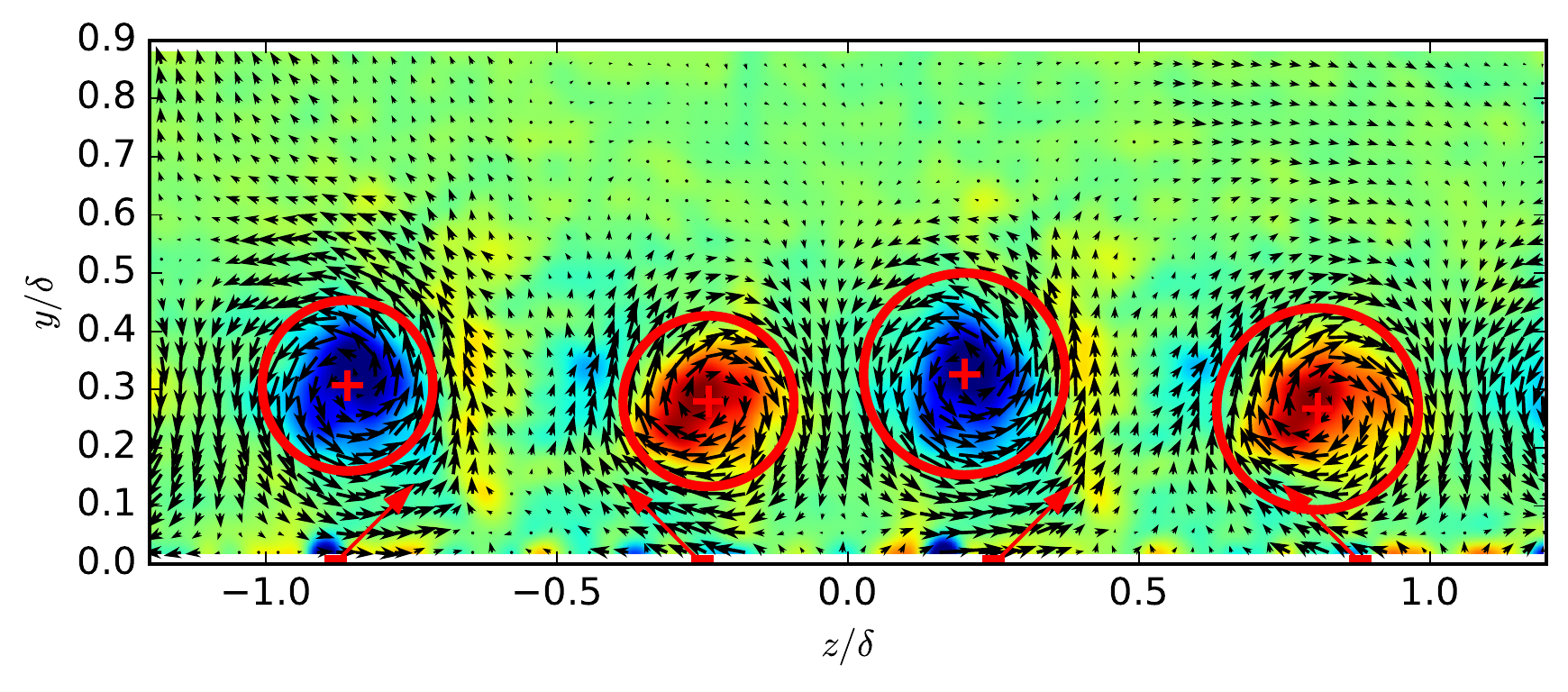} 
\caption{Isocontour of the mean streamwise vorticity $\Omega_x$ ($VR=4$, 100 samples) with the in-plane vector field ($v,w$) superimposed. The extracted radius (red circle) and center of vortices (red crosses) are also superimposed. Red arrows at the wall represents controlled jet locations and orientations.}
\label{fig:mean_detection}
\end{center}
\end{figure}

Figure~\ref{fig:mean_detection} show an example of the vortex detection for the experimental case of $VR=4$. In this figure a center of each vortex is marked by a cross and a circle surrounding it denotes the size of a vortex. 

We transform the Cartesian grid to polar coordinates for each vortex separately, using its center and the radius, arriving at  the tangential velocity $u_\theta$ and the radial velocity  $u_r$, as well as average streamwise vorticity $\Omega_x$. From this polar information, the last parameter, circulation, can be computed using  the integration:  
\begin{equation}\label{eq:Gamma}
\Gamma=\int_{\theta=0}^{\theta=2\pi}\int_{r=0}^{r=R} \Omega_x(r,\theta)\, r dr\,d\theta
\end{equation}




\begin{figure}[htbp]
\begin{center}
\begin{tabular}{ll}
\includegraphics[width=\textwidth]{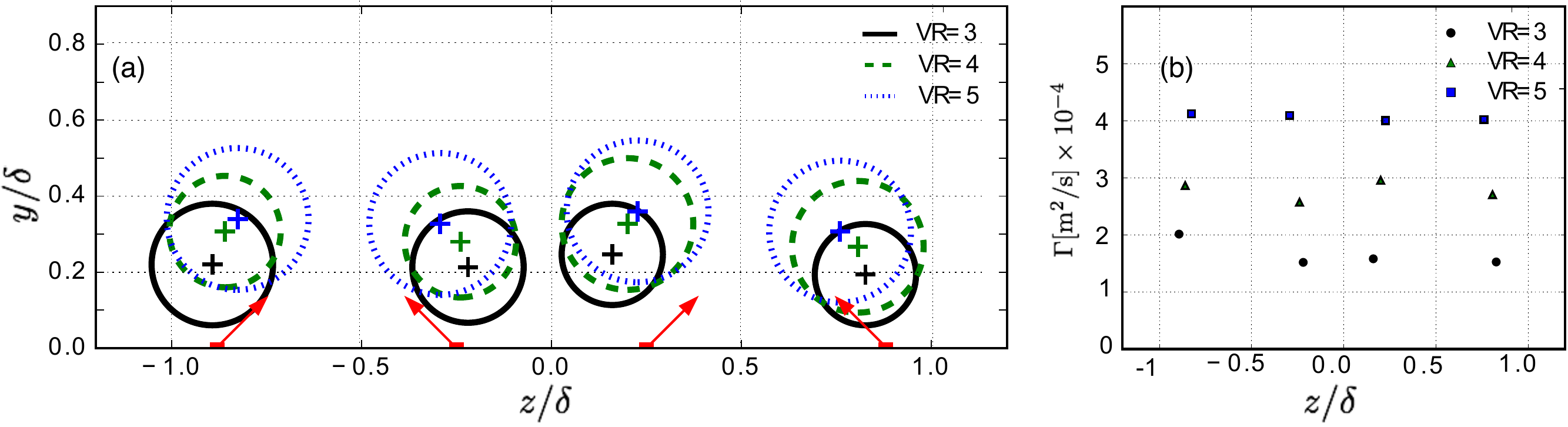}
\end{tabular}
\caption{Effect of $VR$ on vortex properties. (a) Location and radius of each produced streamwise vortex. Red arrows at the wall represents controlled jet locations and orientations. (b) Circulation of the vortices, $\Gamma$ versus their horizontal position, $z/\delta$. }
\label{fig:effetVR}
\end{center}
\end{figure}


Results in figure~\ref{fig:effetVR} show how vortex properties ($y_0,z_0,R,\Gamma$) vary with the control parameter. Produced vortices are displaced further from the wall with increasing $VR$. To compensate for this effect, actuators can blow at different angles from the wall for increasing $VR$. In the future developments, the angle of the jet would become an additional control parameter. The radius of the produced vortices in figure~\ref{fig:effetVR}a, as well as the circulation in figure~\ref{fig:effetVR}b, are shown  to increase with $VR$.

\clearpage
\subsection{Instantaneous vortex identification}

\begin{figure}[htbp]
\begin{tabular}{c}
\includegraphics[width=0.9\textwidth]{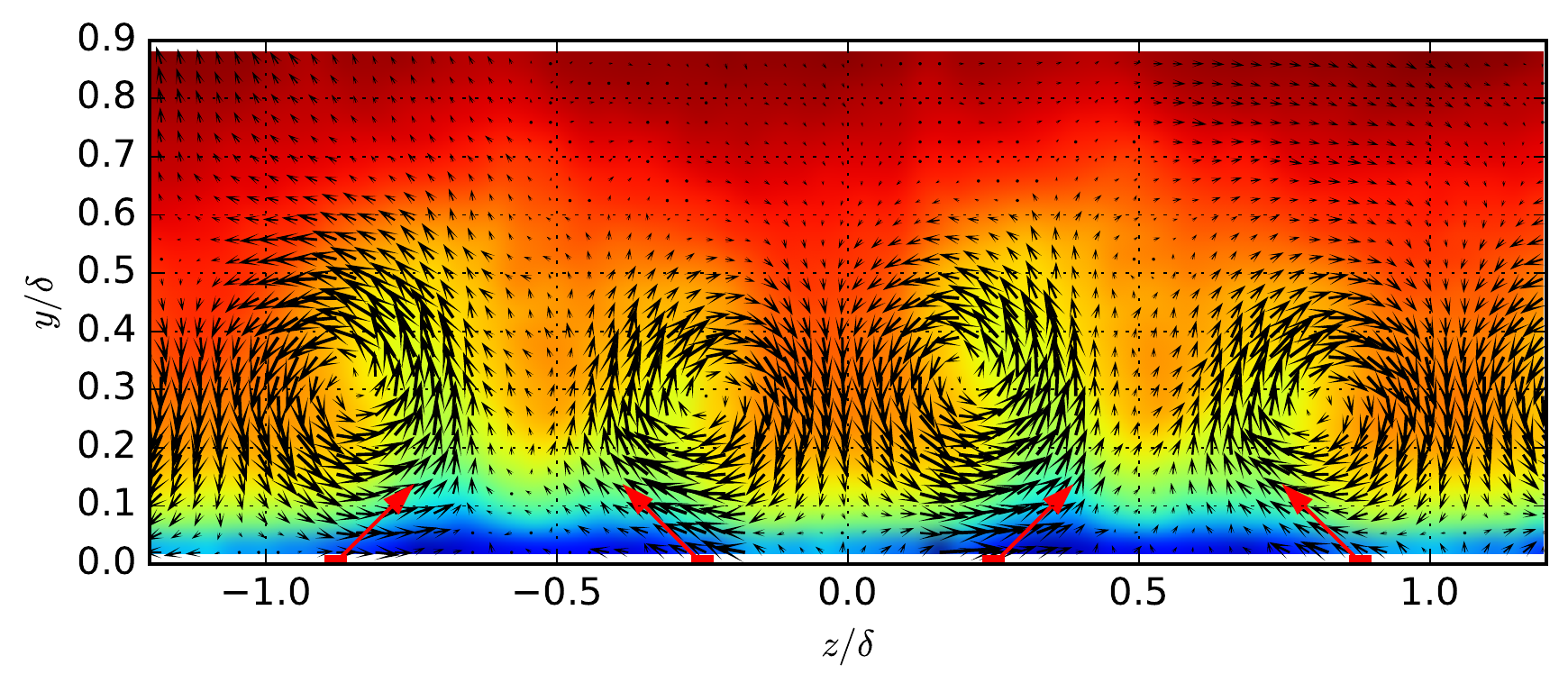}\\
\includegraphics[width=0.9\textwidth]{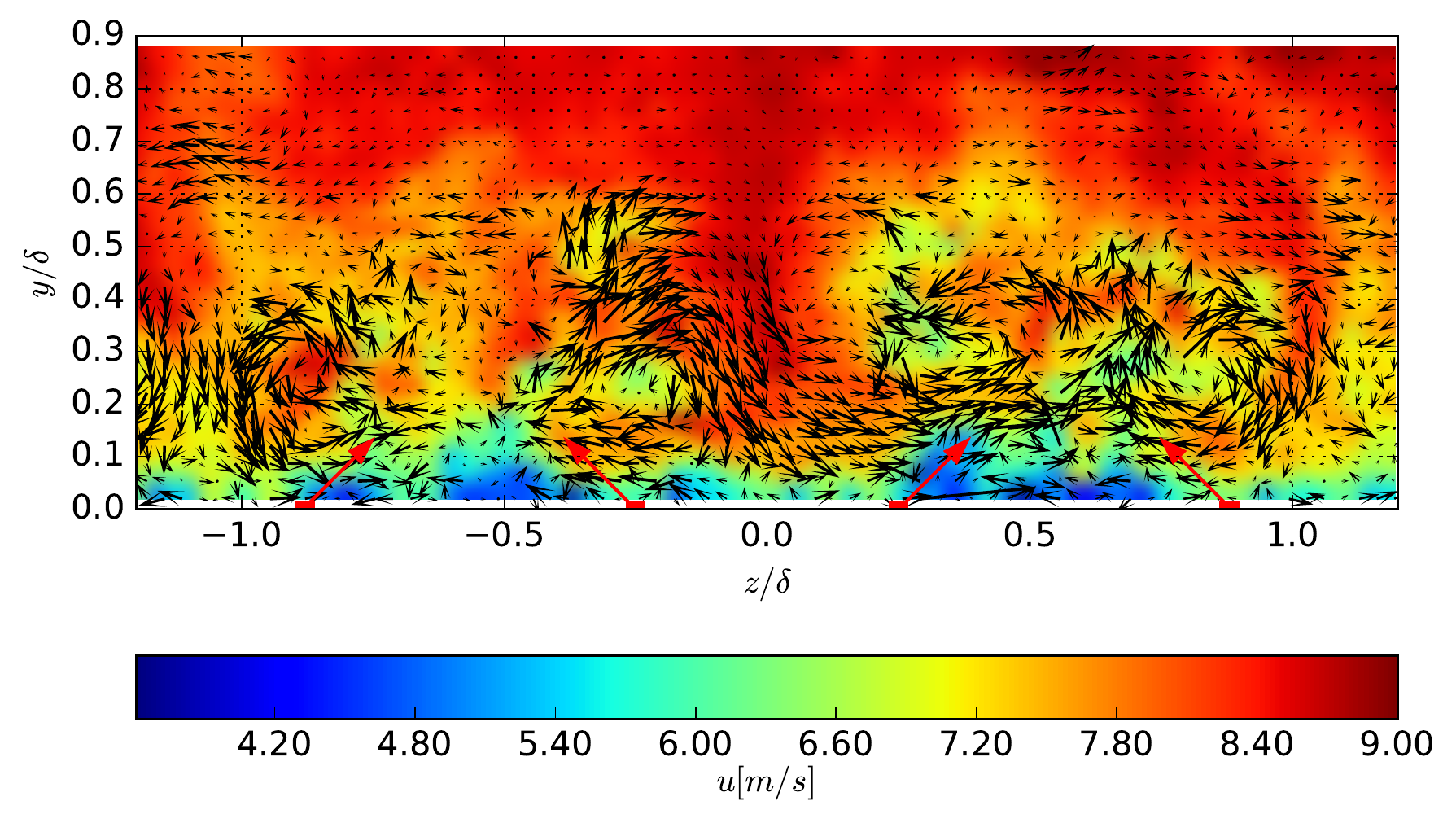}
\end{tabular}
\caption{Isocontours of the streamwise velocity $u$ ($VR=4$, 100 samples) with the in-plane vector field ($v,w$) superimposed. Red arrows at the wall represents controlled jet locations and orientations. (top) the mean field and (bottom) the instantaneous field.}
\label{fig:inst_vortex}
\end{figure}

In order to fulfill the closed-loop feedback-based control strategy, a real-time sensing of the produced vortex properties is needed. Figure~\ref{fig:inst_vortex} exemplifies the major differences between the mean and the instantaneous vector fields, for one case from the dataset of $VR=4$. Differences between mean and instantaneous fields are due to the interaction of the control jet with the turbulent boundary layer flow. 

The first challenge for implementation of the present closed-loop strategy is to identify the instantaneous vortices, using the method developed for an idealized vortex case in the mean flow field. First solution is to apply pre-processing steps that improve the developed identification algorithm: a) trim edges of the PIV field  that are generally much more noisy than the rest of the field due to loss of particles between the two PIV image pairs, b) apply median filter on each velocity component to reduce the background noise in the instantaneous images and c) apply a Gaussian filter prior to localization of the centers of vortices, smoothing of $v(y,z)$ for the vortex center refinement. 

\begin{figure}[htbp]

\begin{picture}(100,200)
\put(0,100){\includegraphics[width=0.9\textwidth]{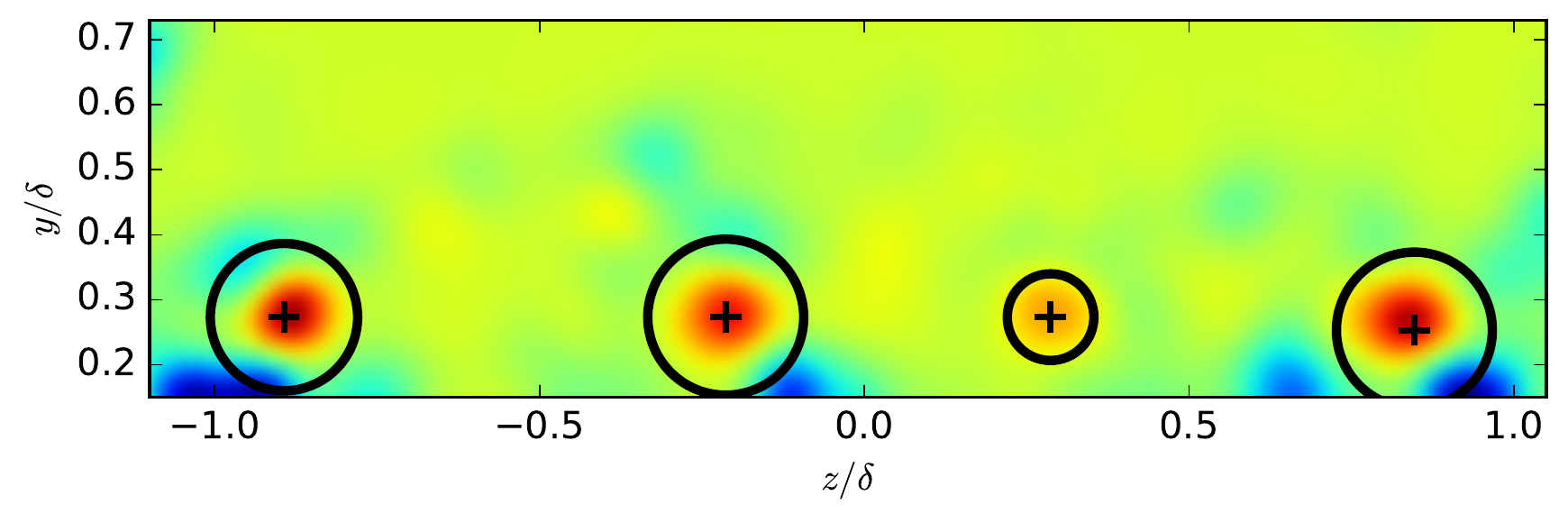}}
\put(35,190){(a)}
\put(0,0){\includegraphics[width=0.9\textwidth]{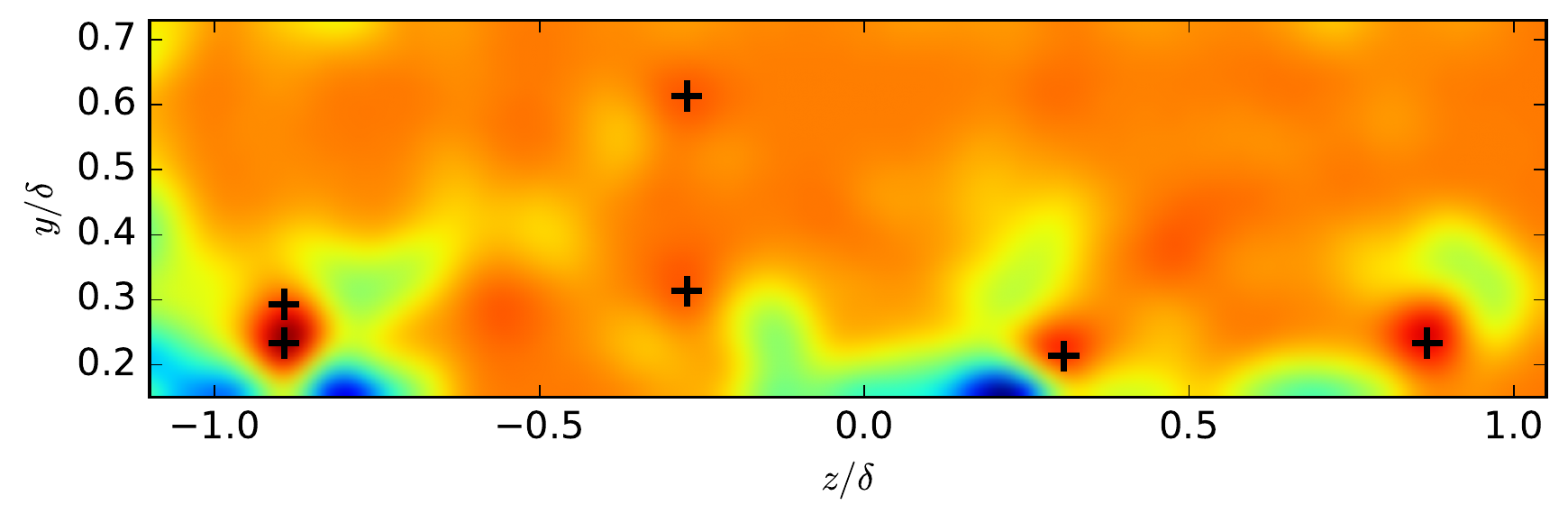}}
\put(35,90){(b)}
\end{picture}
\caption{Examples of application of $Q$ criteria-based identification algorithm to two instantaneous fields with crosses that indicate maxima locations from two examples: a) correct identification example from a set of $VR=4$, b) unsuccessful example from a set of $VR=3$. }
\label{fig:fail_detect}
\end{figure}

We are able to successfully identify vortex properties in $90\%$ of instantaneous samples (see an example in figure \ref{fig:fail_detect}a). Unfortunately, instantaneous fields do not always provide 4 extractable vortices. Figure~\ref{fig:fail_detect}b shows an example of such a case of erroneous detection. Thus, around $z/\delta=-1$, two peaks of the $Q$ criteria are detected instead of a single vortex, around $z/\delta=-0.3$, the $Q$ criteria does not exhibit a peak sufficiently strong relatively to the background noise to be detected. Therefore we need to adjust the robust identification algorithm for an instantaneous detection of vortices in the present state in one of the possible ways: a) apply moving averaging which acts as a low-pass filter in time, or b) adjust the control strategy to work with noisy and wrong identifications.

\begin{figure}[htbp]
\begin{center}
\begin{tabular}{ll}
\includegraphics[width=\textwidth]{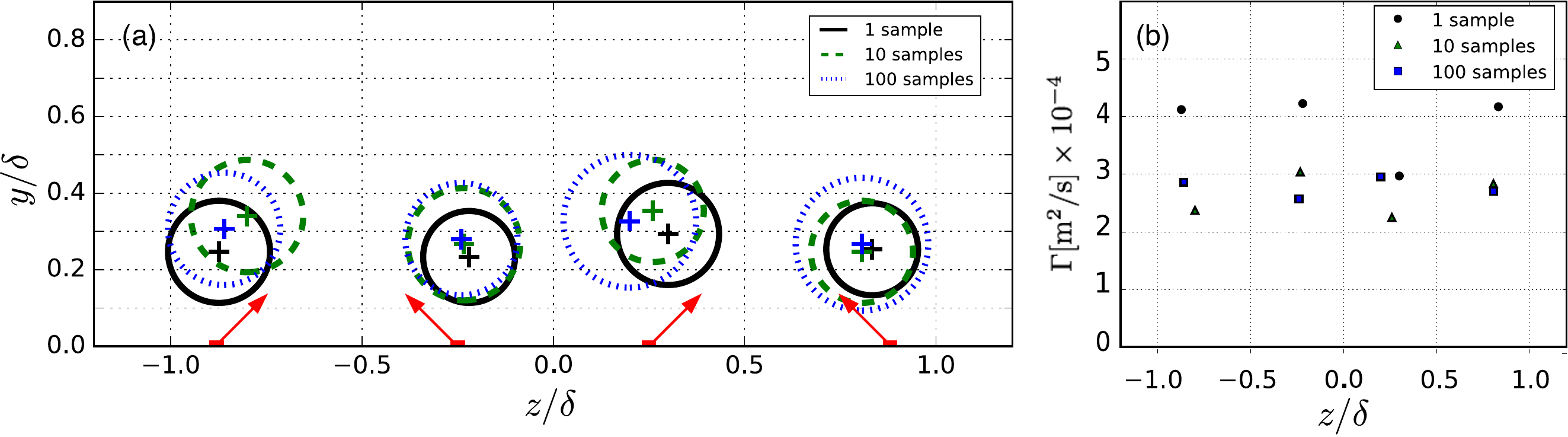}
\end{tabular}
\end{center}
\caption{Average effect on the detected vortex properties, using different number of samples: (a) location and radius, (b) circulation. }
\label{fig:sample_effect}
\end{figure}

We demonstrate the use of a moving window average of a small number of instantaneous PIV realizations. The time scale of the moving window shall be shorter than the closed-loop time scale. In the present dataset we find that 10 samples moving window average provides sufficiently robust vortex detection. Such a detection is presented versus the instantaneous and mean detection (obtained from $100$ samples) in figure~\ref{fig:sample_effect}. Figure~\ref{fig:sample_effect}a shows the detected positions and radii of the vortices for the increasing number of samples. These parameters change in a non-linear fashion with the number of samples and not equally for the four vortices. Circulation of the vortices is shown in figure~\ref{fig:sample_effect}b to vary strongly with instantaneous samples, yet this variation decreases with the increase of the number of samples. We also observe that the values of circulation for the moving window averaging is approaching the mean vortex circulation properties. The small circulation amplitude differences do not matter for closed-loop purposes because the calibration of the associated lift increase and control parameters will be performed with the same moving window averaging and the same number of samples. 






\section{Fast detection}\label{sec:fast_detect}

Real time closed loop control might require even faster identification methods, that on one hand are based on spatial velocity information, but on the other hand provide shorter sensing time. The robust identification method defined in figure~\ref{cmpd:acid}, although successfully identifies the vortices, is based on velocity gradients thus requires 2D spatially resolved velocity field. With a purpose of reducing the computation time, we tried to reduce the spatial resolution required for the velocity gradients field, however every compromise of the spatial resolution reduced significantly the quality of identification. We arrived at another, new identification method, seemingly most suitable for parallelization and real-time computation. The central concept of this method is to avoid computation of spatial velocity gradients for PIV realizations. This step is both the most time consuming and a source of numerical noise. Studying in details the flow field in the given flow field, we came up with the option to use only horizontal (spanwise) vertical velocity component profiles, $v(z)$, presented in the refinement procedure in figure~\ref{fig:refinement}.

\subsection{A ``line to line'' PIV computation} 

Because streamwise vortices are of a finite size which is smaller than the thickness of the boundary layer and because the vortices are located close to the wall, we can save the computation time by reducing the spatial resolution and processing PIV along horizontal lines and with large vertical gaps. The principle is demonstrated using horizontal profiles of the vertical velocity $v(z)$ crossing the vortex center of a simple vortex with a major vorticity component in the out-of-plane direction from PIV Challenge data \cite{piv_challenge}. It is analyzed using OpenPIV \cite{openpiv} as shown below in figure~\ref{fig:challenge_vortex}. 

\begin{figure}[!ht]
\begin{center}
\includegraphics[width=\textwidth]{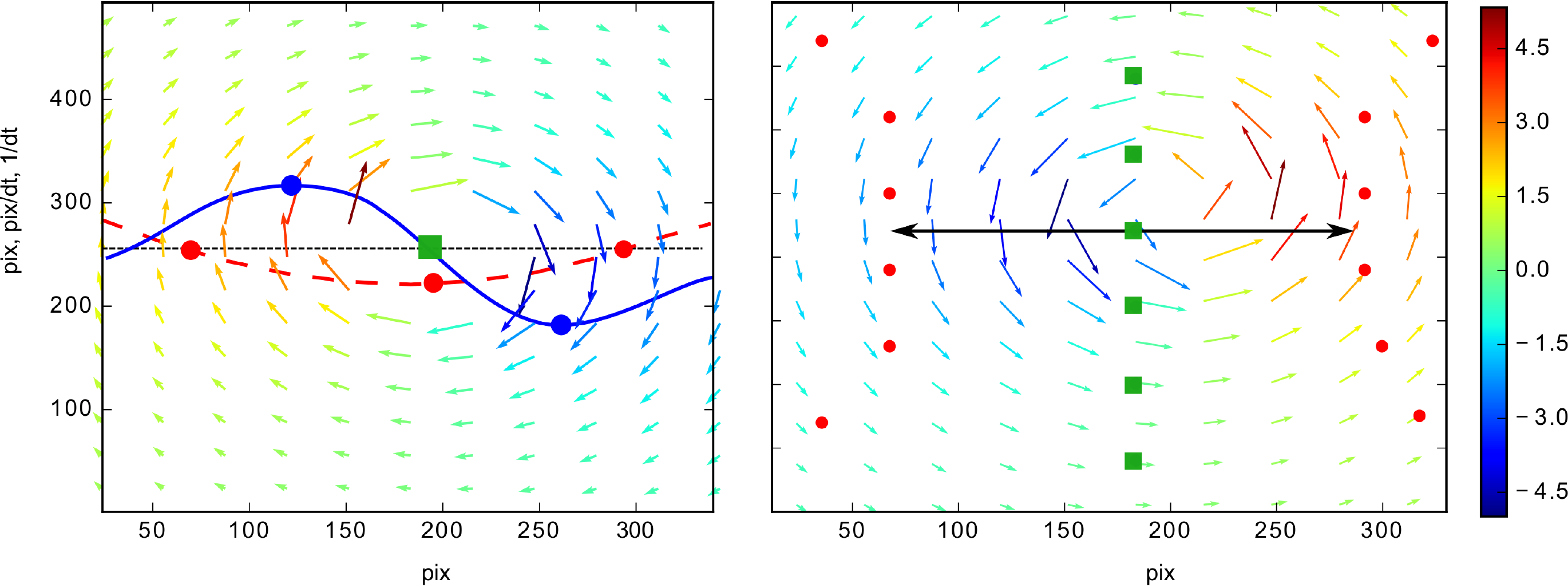}\\
(a) \hspace{5cm} (b)
\end{center}
\caption{Demonstration of the vertical velocity identification method, using vector field of a vortex from a PIV Challenge test case. (a) Arrows denote the vector field, color of the arrows and the color scale correspond to the magnitude of vertical component, $v(y,z)$. Solid (blue) line is a horizontal velocity profile $v(z = z_0)$ at the line that corresponds to the vortex center, dashed (red) curve is a smoothed horizontal derivative $\partial v/ \partial z(z = z_0)$. Dots mark the extrema and the zero crossings of $v(z)$ or of the derivative $\partial v/ \partial z$. (b) Symbols marking the zero crossing and the extrema position at several horizontal positions below and above $z_0$. The horizontal size of a vortex is marked by the black line at the position where the distance between the crossing points is minimal. \label{fig:challenge_vortex}}
\end{figure}
 
The identification method is straightforward for the presented type of streamwise vortices measured in the transverse -- wall-normal plane. The vertical velocity profile $v(z)$ crossing a vortex is a double-peak curve, crossing zero at a vortex center, as shown by a solid (blue) curve in figure~\ref{fig:challenge_vortex} on top of which dots mark the peaks and zero crossings. The derivative of the horizontal profile $\partial v/\partial z$, its extrema and the two neighbor zero-crossing points are shown by a dashed (red) curve in  figure~\ref{fig:challenge_vortex} and also marked by dots. To summarize, the ad-hoc method that allows for simple and robust vortex identification can use: 
\begin{enumerate}
\item PIV analysis of a small region of interest, for instance a row of 32 pixels in height for the full width of the PIV image, starting from the bottom of the image (in our case corresponding to the wall surface); 
\item Velocity profile $v(z)$ and the derivative $\partial v/\partial z$ of the vertical velocity component and search for peaks and zero crossing points of a single or both curves. 
\item If vortices have been located, shift the region of interest to the next row (in the upward direction) at smaller step, otherwise the step is approximately an average size of the vortex at this given streamwise position. 
\item The search stops when distance between zero-crossings changes sign from decreasing to increasing, as seen in  figure~\ref{fig:challenge_vortex}b.
\end{enumerate}

The time consumption of this analysis is substantially smaller compared to the $Q$-criterion identification method. It is possible to further reduce the computation time using a 1D FFT analysis in vertical direction instead of two 1D FFT runs used for the full PIV vector analysis. Furthermore, we could run the method using only a pixel-displacement version, saving time on sub-pixel peak search. The method is not as accurate as the robust identification method, but it could be a solution in a view of the present computation consumption of the vortex identification methods based on vorticity and circulation. 

\subsection{Results of $v(z)$ vortex identification} 

\begin{figure}[!ht]
\centering\includegraphics[width=.9\textwidth]{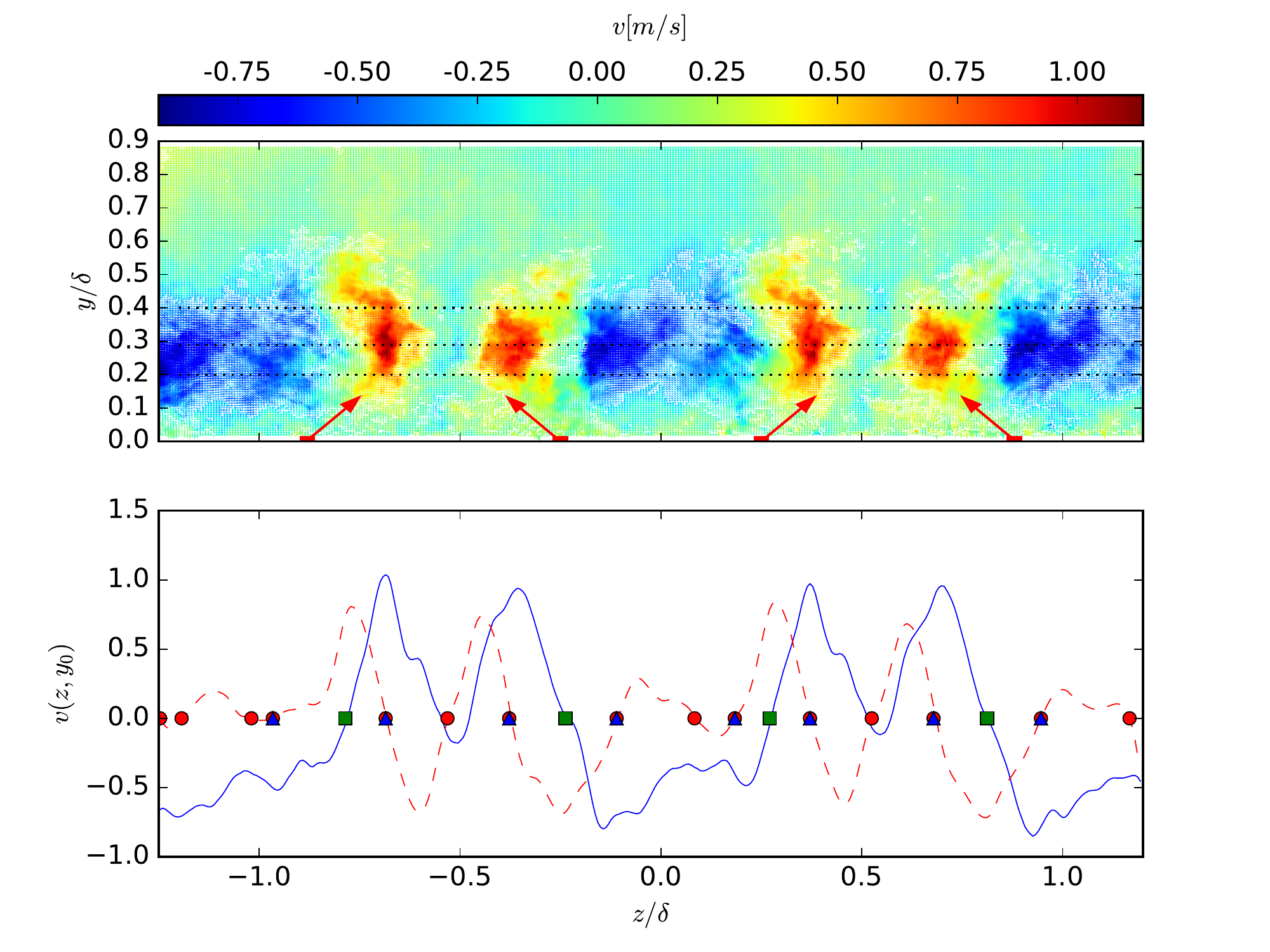}
\caption{(top) Instantaneous velocity field for $VR=4$, averaged over 10 samples, color map corresponds to the vertical velocity values. (bottom) Velocity profile $v(z)$ (solid curve) and its horizontal derivative $\partial v/\partial z$ (dashed curve) for the line of $z=0.3$, an example of the vortex detection using the horizontal profile. Green squared, red dots and Blue triangular mark respectively vortex centers, zero crossings and zero-crossings around the vortex centers using the ''line to line" method.}\label{fig:line_method}
\end{figure}

\begin{figure}[htbp]
\begin{tabular}{ll}
\includegraphics[width=.5\textwidth]{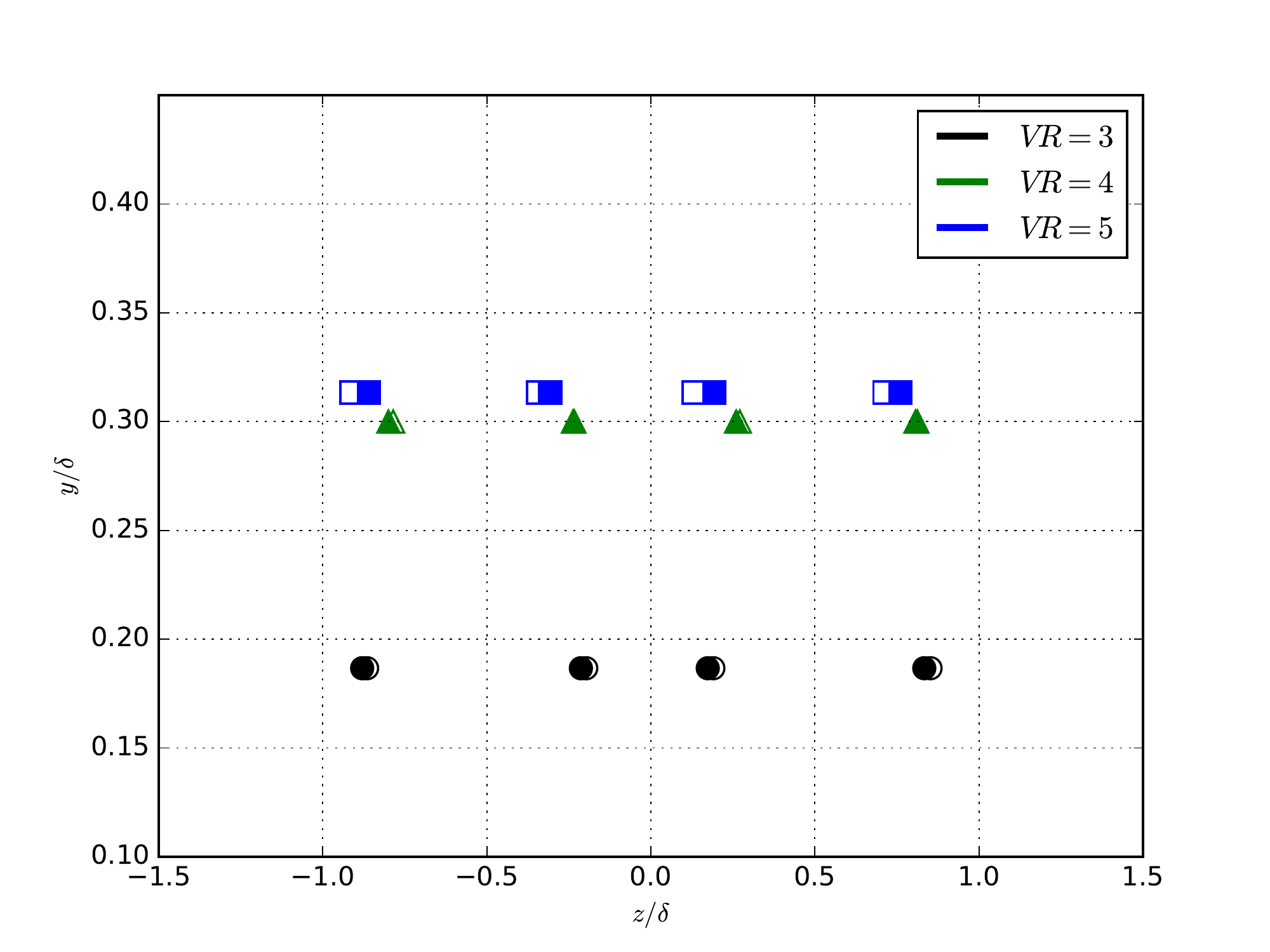} & \includegraphics[width=.48\textwidth]{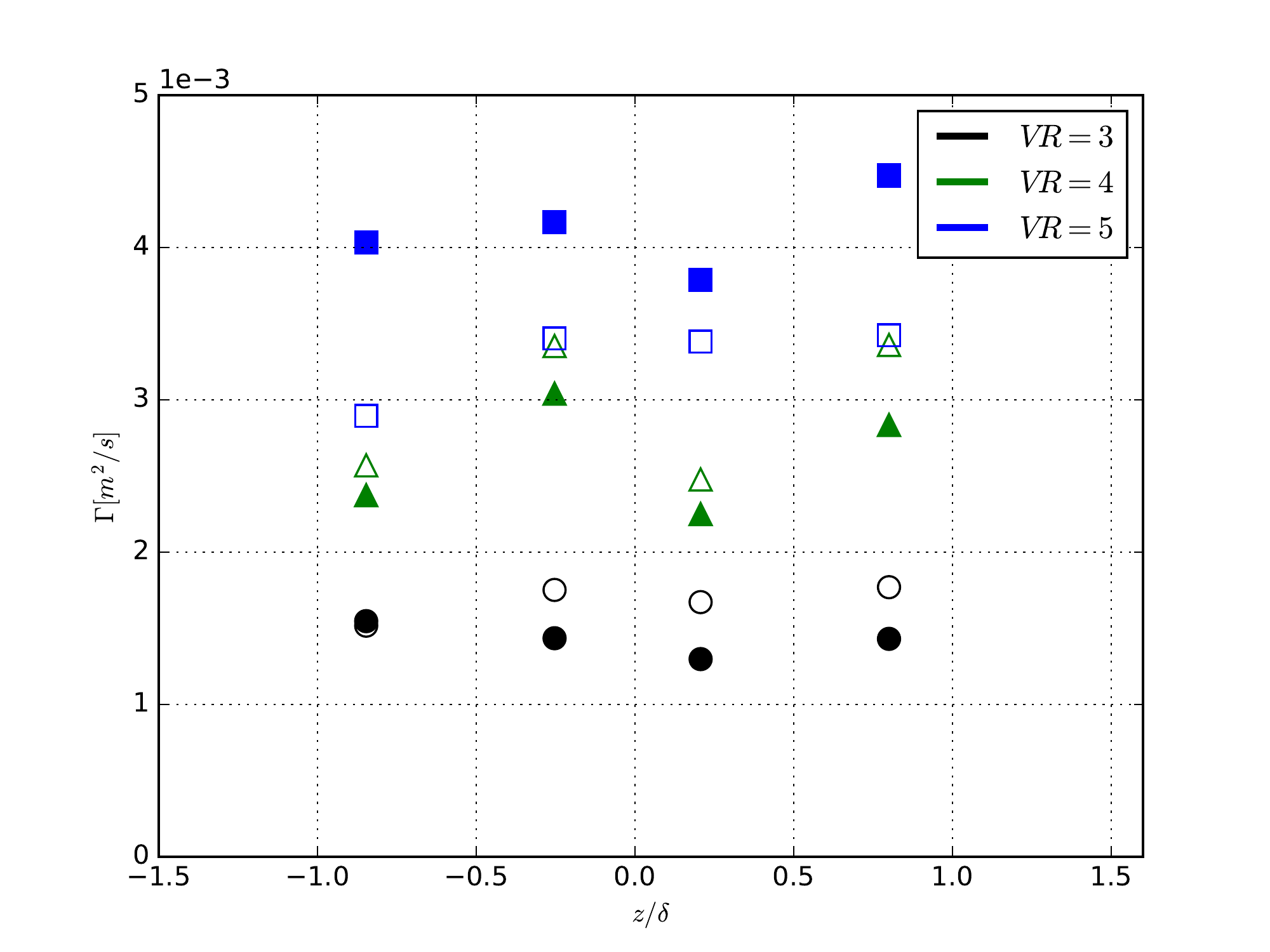}\\
(a) & (b)
\end{tabular}
\caption{Comparison between line (unfilled symbols) and robust (filled symbols) detection methods: (a) centers and (b) $10\Gamma$ compared to $\Gamma_y$.}
\label{fig:comp_detect}
\end{figure}

It is hypothesized that the vertical velocity profile at the vortex center contains all informations needed to characterize the circulation used in closed-loop control strategies. For sake of comparison, we use the moving window (10 samples) averaged PIV velocity fields produced behind active vortex generators for $VR=4$,  as shown in top panel of Figure~\ref{fig:line_method}. The colormap is of the vertical velocity, $v(y,z)$ with blue regions marking downwash and red regions upwash. The bottom panel we demonstrate the profile of $v(z)$ (solid curve) and its horizontal derivative, $\partial v/\partial z$ (dashed curve), approximately through the central line of the four vortices. Each vortex is a double-peak curve, crossing zero at a vortex center, as shown in figure~\ref{fig:line_method}. Although the method could use the velocity directly, identifying a template of half a period of a sine curve or a Gaussian distribution curve, a more robust detector of a center is a smoothed velocity derivative. In this case the algorithm looks for an extrema and two neighbor zero-crossing points of a $\partial v/\partial z$, as shown by a dashed (red) curve in  figure~\ref{fig:line_method} and marked by symbols. The size of the vortex is then the distance between the zero crossing points (marked blue triangular). Obviously for the clockwise vortex the definitions are similar and mirrored, thus we search for the maxima of the derivative instead of a minima.
If the number of maxima is higher than four, one should thus keep the four strongest vortices using the magnitude of the extrema of $v$. For robustness, the profiles at different heights (in this particular case $y/\delta$ between 0.2 to 0.4) can be used to detect center of vortices. Results of this line detection is compared to results from the robust detection for three $VR=3,4,5$ (see figure \ref{fig:comp_detect}a). The detection of vortex centers is identical for both methods used. 


Circulation can be approximated using the integral of $\partial v/\partial z$ along the horizontal line crossing the vortex center. The vortex circulation, $\Gamma$, is defined as follows:
\begin{equation}\label{eq:gamma}
\Gamma=\int_{S_{vort}}  \left(\frac{\partial w}{\partial y}-\frac{\partial v}{\partial z}\right) dS_{vort}  
\end{equation}
\noindent with $S_{vort}$ the surface delimited by the vortex radius. For the horizontal lines crossing the vortex center $\Delta w$ is close to zero and the strength of the vortex can be approximated using the component $\Gamma_y$: 
\begin{equation}\label{eq:reducedGamma}
\Gamma(\Delta y)=\Gamma_y\simeq -\int_{z_0-R}^{z_0+R}\frac{\partial v(z,y_0)}{\partial z}\, dz
\end{equation}
\noindent with $v(z,y_0)$ the line profile of the vertical velocity crossing the vortex center and end points for integration on the left/right side of the vortex or $z_0 \pm R$. 

We compare the approximate circulation to the one computed in the robust method in figure~\ref{fig:comp_detect}b.  A good correspondence between $|\Gamma_y|$ from the fast method and $|\Gamma|$ from the robust identification is found for $VR=3,4$ as shown in figure \ref{fig:comp_detect}b. This proportionality seems however to be changed for the third case, $VR=5$, with the $|\Gamma_y|$ that is only slightly increased with $VR$, contrary to what is found in the robust detection of the circulation (see figure~\ref{fig:effetVR}b). We attribute this effect to the appearance of the vortices in figure \ref{fig:effetVR}: a) vortices are distorted and less axisymmetric, and b) larger vortices appear closer to each other and affect the identification of the zero-crossings. 
%
%
To improve the fast ``line'' detection, a contribution in $y$ direction could be easily added using also the vertical profile:
\begin{equation}\label{eq:Gz}
\Gamma(\Delta z)=\Gamma_z\simeq -\int_{y_0-R}^{y_0+R}\frac{\partial w(z_0,y)}{\partial y\, dy}
\end{equation}
\noindent with $w(z_0,y)$ the vertical profile of the horizontal velocity component crossing the vortex center. However, for the sake of the present demonstration, it is sufficient to note that for cases of close vortices, the line detection method could include more than one direction as in Eqs.~\eqref{eq:reducedGamma} and Eq.~\eqref{eq:Gz} in order to improve robustness. 


\section{Control strategy implementation }
\label{sec:control_strategy_implementation}

\begin{figure}[htbp]
\begin{center}
  \includegraphics[width=.8\textwidth]{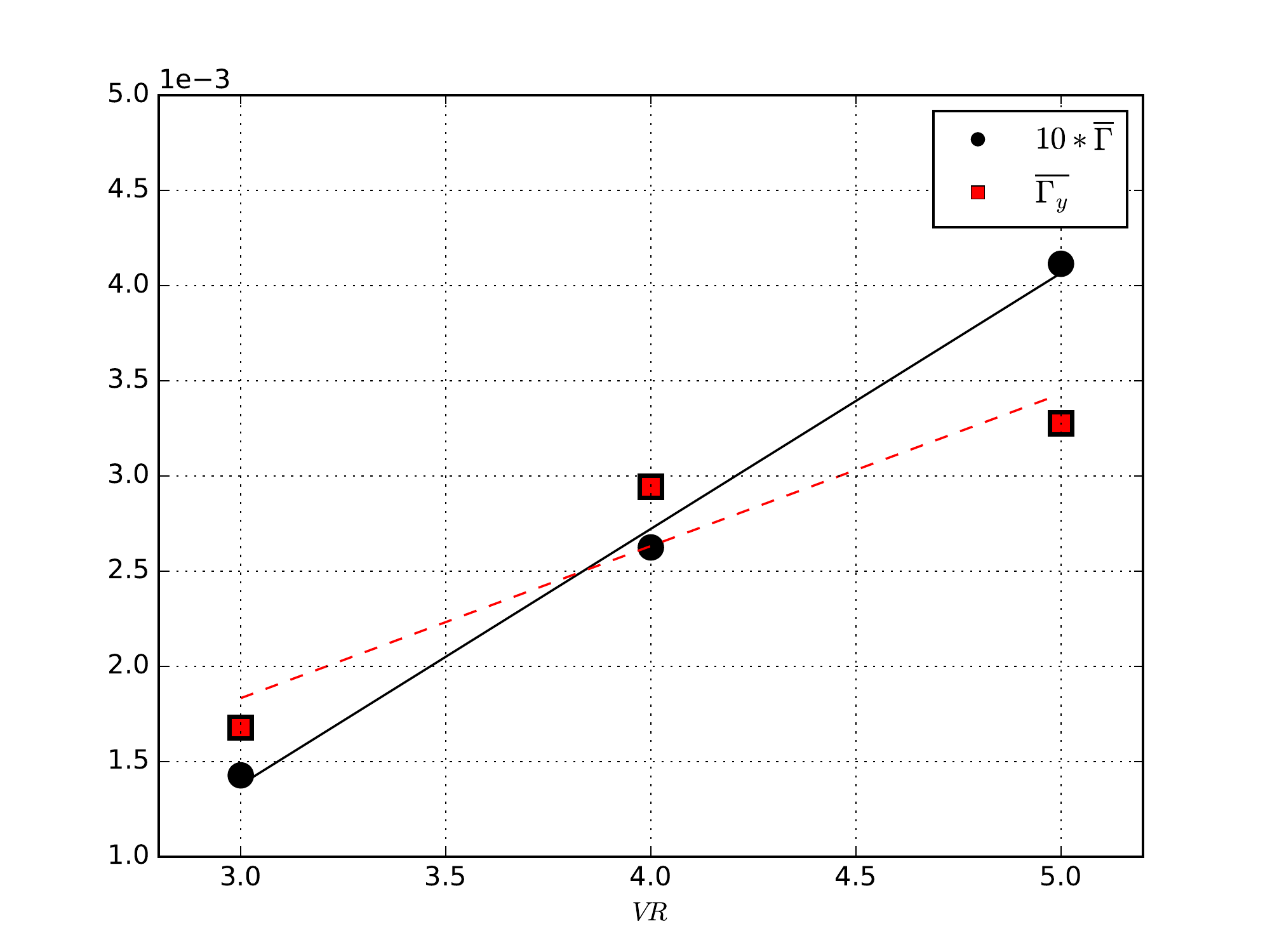}
\caption{Strength (from line detection $\overline{|\Gamma_y|}$ or robust detection $\overline{|\Gamma|}$) of the produced streamwise vortices averaged over the four vortices in z direction, versus the control parameter. This represent the steady state response of the system to control. }
\end{center}
\label{fig:steady_state}
\end{figure}

For wind turbine applications, a simple control strategy can be proposed from previous results using the average of $|\Gamma_y|$ over the four vortices $\overline{|\Gamma_y|}$ for robustness. The implementation can be described from the  block diagram shown in figure\ref{cmpd:control} with  $\Gamma_r=\overline{|\Gamma_y|}$ from the line identification. The desired circulation $\Gamma_o$ is given from the off-line steady state relationship using the robust control identification,   $\overline{|\Gamma|}\simeq \overline{\Gamma_y}/10$ (see figure \ref{fig:steady_state}). What remains now is to study the transient states of the closed loop. Note that this control system can be more elaborate with a desired circulation distribution in $z$ direction, that can be driven from upstream measurements of  the turbulent boundary layer properties such as high speed or low speed streaks for instance~\cite{Ganapathisubramani_al2003}.

\section{Conclusions}\label{sec:conclusions}

In this study we propose the robust and fast vortex characterization methods for the purpose of a real-time (online or on-the-fly) feedback-based control, based on active vortex generators and PIV sensing approach.  Both methods can extract centers and strengths of streamwise vortices generated behind active vortex generators in a turbulent boundary layer flow, and we show how to integrate those in closed-loop control strategies. For the feasibility study we use a previously obtained dataset of SPIV measurements in the transverse-wall-normal plane behind active devices. A robust algorithm is using the $Q$-criteria and the integration of vorticity of each extracted vortex. Results show that a moving window average of a small number of instantaneous fields is nevertheless needed for increased robustness.  Additional, new method is developed to minimize the computational effort and it uses only horizontal lines of vertical velocity with the potential to significantly cut down the computational effort related to full scale PIV computation followed by spatial derivatives. It has been applied to the available dataset, compared to the robust method and a following control strategy is proposed. Future work will propose an implementation of this strategy using software or hardware-based computations of PIV and a feedback-based operation of active vortex generators.

\bibliographystyle{plain}
\bibliography{bib}

\begin{thebibliography}{10}

\bibitem{becker2007}
R.~Becker, R.~King, R.~Petz, and W.~Nitsche.
\newblock Adative closed-loop separation control on a high-lift configuration
  using extremum seeking.
\newblock {\em AIAA Journal}, 45(6):1382--1392, 2007.

\bibitem{carlier2005}
J.~Carlier and M.~Stanislas.
\newblock Experimental study of eddy structures in a turbulent boundary layer
  using particle image velocimetry.
\newblock {\em J. Fluid Mech.}, 535:143--188, 2005.

\bibitem{Chakraborty_al2005}
P.~Chakraborty, S.~Balachandar, and R.J. Adrian.
\newblock On the relationships between local vortex identification schemes.
\newblock {\em J. Fluid Mech.}, 535:189--214, 2005.

\bibitem{foucaut_al2014}
J.M. Foucaut, S.~Coudert, C.~Braud, and C.~Velte.
\newblock Influence of a light sheet separation on the spiv measurement in a
  large field spanwise plane.
\newblock {\em Meas. Sci. Technol.}, 25:035304, 2014.

\bibitem{Ganapathisubramani_al2003}
B.~Ganapathisubramani, E.K. Longmire, and I.~Marusic.
\newblock Characteristics of vortex packets in turbulent boundary layers.
\newblock {\em Journal of Fluid Mechanics}, 478:35--46, 2003.

\bibitem{gautier_aider2013}
N.~Gautier and J.-L. Aider.
\newblock Control of the separated flow downstream of a backward-facing step
  using visual feedback.
\newblock {\em Proc. Roy. Soc. A}, 469(2160), 2013.

\bibitem{lin2002}
J.C. Lin.
\newblock Review of research on low-profil vortex generators to control
  boundary layer separation.
\newblock {\em Progress in Aerospace Sciences}, 38:389--420, 2002.

\bibitem{peterson2004}
S.~D. Peterson and M.~W. Plesniak.
\newblock Evolution of jets emanating from short holes into crossflow.
\newblock {\em Journal of fluid mechanics}, 503:57--91, 2004.

\bibitem{shaqarin2013}
T.~Shaqarin, C.~Braud, S.~Coudert, and M.~Stanislas.
\newblock Open and closed-loop experiments to identify the separated flow
  dynamics of a thick tbl.
\newblock {\em Experiments in Fluids}, 54(1448), 2013.

\bibitem{piv_challenge}
M.~Stanislas, K.~Okamoto, and C.J. K{\"a}hler.
\newblock Main results of the first international {PIV} challenge.
\newblock {\em Meas. Sci. Technol.}, 14:R63--R89, 2003.

\bibitem{openpiv}
Z.~J. Taylor, R.~Gurka, G.~A. Kopp, and A.~Liberzon.
\newblock Long-duration time-resolved piv to study unsteady aerodynamics.
\newblock {\em IEEE Trans. Instrum. Meas.}, 59(12):3262--3269, 2010.

\bibitem{tilmann2000}
C.P. Tilmann, K.~L. Langan, J.G. Betterton, and M.J. Wilson.
\newblock Characterization of pulsed vortex generator jets for active flow
  control.
\newblock In {\em RTO AVT Symposium on "Active Control Technology for Enhanced
  Performance Operational Capabilities of Military Aircraft, Land Vehicles and
  Sea Vehicles}, 2000.

\bibitem{willert_al2010}
C.~Willert and M.~G.~M. Munson.
\newblock Real-time particle image velocimetry for closed-loop flow control
  applications.
\newblock In {\em 15th Int. Symp on Appl. Laser Techniques to Fluid Mechanics},
  2010.

\end{thebibliography}

\end{document}